\documentclass{vgtc}                          

\graphicspath{{figures/}{pictures/}{images/}{./}} 

\usepackage{times}                     

\usepackage{tabu}                      
\usepackage{booktabs}                  
\usepackage{lipsum}                    
\usepackage{mwe}                       

\usepackage{mathptmx}                  
\usepackage{xspace}

\onlineid{0}

\vgtccategory{Research}

\vgtcinsertpkg

\title{\scifivis: Way Out There -- How \scifi and Visualization Influence~Each~Other
}

\author{Ulrik Günther\thanks{e-mail: ulrik.guenther@hzdr.de}\\ %
     \parbox{1.4in}{\scriptsize \centering Helmholtz-Zentrum Dresden-Rossendorf e.V., Germany} 
\and Juli\'{a}n M\'{e}ndez\thanks{e-mail: julian.mendez2@tu-dresden.de}\\ %
     \parbox{1.8in}{\scriptsize \centering Interactive Media Lab Dresden\\ TU Dresden, Germany}
\and Gabriela Molina Le\'{o}n\thanks{e-mail: leon@cs.au.dk}\\ %
     \scriptsize Aarhus University, Denmark
\and Samuel Pantze\thanks{e-mail: s.pantze@hzdr.de}\\ %
     \parbox{1.4in}{\scriptsize \centering Center for Advanced Systems Understanding (CASUS), Görlitz, Germany} 
\and Mario Romero\thanks{e-mail: mario.romero@liu.se}\\ %
     \scriptsize Linköping University, Sweden %
\and Abdulhaq Adetunji Salako\thanks{e-mail: abdulhaq.salako@uni-rostock.de}\\ %
    \parbox{1.4in}{\scriptsize \centering Institute of Visual and Analytic Computing, University of Rostock, Germany} 
\and Annalena Ulschmid\thanks{e-mail: annalena.ulschmid@tuwien.ac.at}\\ %
     \scriptsize TU Wien, Austria
}

\abstract{We propose a hybrid half-day workshop at IEEE VIS 2026, calling for participation from visualization researchers and science fiction creators in order to develop a systematic understanding of the two-way relationship these communities have long shared. We invite submissions of creative formats showcasing connections and inspiring future research. Our workshop plan includes a keynote, lightning talks, brainstorming, cross-community critique, affinity mapping, and discussion around identified themes.} 


\begin{document}



\newcommand{\scifivis}{SciFi-VIS\xspace}
\newcommand{\scifi}{SciFi\xspace}

\maketitle


\firstsection{Introduction}

\section{Introduction}

    Visualization research and Science Fiction (\scifi) media have a long-standing two-way relationship. \scifi films, series, and games frequently depict data interfaces and analytic workflows that shape public expectations of what visualization and human-computer interaction (HCI) is and what it could be \cite{kiianmies_exploring_2026, krings_what_2023}. At the same time, visualization (VIS) researchers often reference, adopt, or reinterpret interface ideas that appear in well-known fictional worlds \cite{jordan_exploring_2018, blythe_research_2017, kong_neuromancer_2021}.
    Despite this connection, the VIS community is yet to develop a shared, systematic understanding of this relationship regarding:
    \begin{itemize}
    \setlength{\itemsep}{0pt}
        \item What visualization ideas are derived or borrowed from \scifi media? Which concepts have not yet been adapted and what are the reasons (computability, technical feasibility, ...)?
        \item Do researchers credit the source of inspiration? If yes, how?
        \item Do \scifi creators and user interface practitioners actively consult VIS work? Or does inspiration flow primarily through informal channels?
    \end{itemize}

    IEEE VIS workshops are designed to incubate emerging communities and provide an ideal forum for work that is too niche for the main conference tracks. \scifivis directly targets that niche by connecting visualization experts with media creators, interface designers, gaming practitioners, and scholars of speculative design.

\section{Related Prior Workshops and Seminars}

    We propose \scifivis as a new workshop format. To our knowledge, no prior workshop with this scope has taken place at VIS so far. Related panels, workshops, and online resources include:
    \begin{itemize}
    \setlength{\itemsep}{0pt}
        \item A CHI 1992 panel chaired by Aaron Marcus, in which \scifi writers tried to predict future user interfaces \cite{marcus_sci-fi_1992}. He repeated the panel at CHI 1999 \cite{marcus_opening_1999}.
        \item An ACE 2016 workshop, held to create a research agenda about \scifi-inspired HCI research \cite{mubin_towards_2016}, stating that \emph{[...] sprinklings of \scifi discourse in HCI academic venues [...] are deﬁnitely not a regular occurrence. Ideally this could be accomplished as a ﬁxed session or track at an annual ﬂagship HCI conference; where both HCI researchers and \scifi authors will be invited to present ﬁctional visions of their work from both domains.}
        \item A VIS 2020 \emph{VisFutures} workshop to envision futuristic design tools \cite{noauthor_ieee_nodate}, which did not focus on existing \scifi work
        \item Two workshops on  the topic \emph{HCI/UX in Science-Fiction Movies/TV: What Can We Learn from the Past 100 Years of the Future?}, conducted at HCII 2021 \cite{noauthor_t15_nodate} and HCII 2022 \cite{noauthor_t12_nodate}, respectively, in which the authors outline the evolution of \scifi over the last century, and analyze existing \scifi movies and TV shows with regard to HCI and UX design
        \item The blog website \texttt{\href{https://scifiinterfaces.com}{scifiinterfaces.com}} by Christopher Noessel, which provides extensive analyses of a vast corpus of \scifi movies and TV shows
    \end{itemize}
    
    While these events and resources have provided opportunities to discuss intersections between HCI work and \scifi, our workshop will be the first event to focus on the intersection of data visualization and science fiction, aiming to also connect with \scifi creators and designers. Most of these related events and resources focused on either only movies and TV shows, or novels---which cannot provide a robust foundation for disseminating visualizations by their very nature. In contrast, our workshop will also incorporate visualizations from computer games.

\section{Workshop Goals and Research Questions}

    The VIS conference has strong research threads on interaction, immersive analytics, storytelling, and design. \scifivis complements those areas by focusing on the origin stories of visualization work inspired by \scifi and on the cultural pipeline that moves ideas between \scifi and visualization work.
    Through a workshop format, the community will benefit from open discussions, hands-on collaborative annotation, and synthesis rather than one-way presentations.
    
    We aim to explore how \scifi media inspires visualization research and design. We will invite submissions that investigate instances where \scifi has influenced real-world visualization
    systems, analyzing successful examples of this collaboration, and the benefits of explicitly citing these inspirations.
    Thus, we welcome submissions illustrating or answering questions such as:
    \begin{itemize}
    \setlength{\itemsep}{0pt}
        \item What inspiration have researchers received from \scifi when designing their visualization interfaces? How do creators decide on interfaces as plot devices, and what can we learn from their creative
        process? 
        \item How are decisions for \scifi interfaces made and what informs these decisions? 
        \item What are underutilized concepts from \scifi media that we may employ to design visualization interfaces and vice-versa? Are there concepts that remain infeasible (``way out there'') to translate at present, and what factors account for this infeasibility?
    \end{itemize}
    
   We are interested in building bridges between \scifi creators and consumers, and visualization researchers and practitioners. In the workshop, we will encourage participants to examine the feasibility and usability of \scifi interfaces and to look for design patterns that might inspire future visualization work and give us insights into potentially underutilized concepts that may already be employed either in the VIS or the \scifi community, but not in the other.

\section{Format and Planned Activities}

    \scifivis is designed as a \textbf{half-day} workshop. We will use a keynote and lightning talks to seed discussion, followed by structured collaborative activities that produce tangible outputs. We plan to engage participants in hands-on creative activities, potentially involving whiteboards, sketch pads, or AI image generators to brainstorm about interfaces and interaction modalities that push the boundaries
    of current \scifi tropes.
    
    We will reach out and invite practitioners in the film, television, and game industry who focus on interface design, as well as visualization researchers working on immersive analytics, interaction design, speculative design, or media studies. For a broader participation, we aim to
    conduct the workshop in a hybrid manner.
    To give attendees time to collect, discuss, and brainstorm ideas, we propose the following group activities and schedule:
    
        \begin{enumerate}
            \item\textbf{Welcome \& framing} (15 minutes). Organizers present goals, norms, schedule, and a quick overview of expected outputs. 
            \item\textbf{Keynote} (30 minutes), including audience questions.
            \item\textbf{Lightning talks} (45 minutes). Up to 9 selected talks of 5 minutes each, to be adjusted according to accepted submissions. 
            \item\textbf{Poster/demo session} (30 minutes), extendable into the break. 
            \item\textbf{Break} (15 minutes).
            \item\textbf{Group-based brainstorming} (30 minutes). Participants design and conceptualize novel visualization interfaces as a hands-on activity, inspired by design fiction methods~\cite{noauthor_ieee_nodate}. The results will be presented and discussed.
            \item\textbf{Cross-community critique} (25 minutes), showcasing the results of the brainstorming session, plus general examination of examples from visualization and \scifi work in both directions, \scifi to research, and research to production, identifying mismatches and opportunities.
            \item\textbf{Affinity mapping} (30 minutes). Groups build affinity maps (clustering annotations into themes) and propose taxonomy dimensions (e.g., interaction metaphor, narrative function, plausibility cues). 
            \item\textbf{Discussion, closing synthesis and next steps} (30 minutes). Participants discuss citation practices, responsible inspiration, and a potential research agenda. Vote on top action items and assign small volunteer teams for post-workshop outputs.
    \end{enumerate}
    
    We will adapt the final schedule to the conference constraints.

\section{Accommodations and Contingencies}

    We will require standard audiovisual equipment, meaning a projector or screen, with microphones and audio playback for presenting, and a stable internet connection, to accommodate hybrid participation. We would appreciate a cabaret-style or round-table room setup to support working in smaller groups of four to six people.

\section{Publication Plan}

    We are planning \scifivis as a non-archival workshop to increase chances of participating with less formal submission styles. Position statements and posters can be hosted (if the authors consent) on the workshop website and an open repository.

\section{Call for Participation and Timeline}

    We are looking for more than just the usual research papers. Instead, attendees are invited to submit creative and non-traditional works that show off the connection between \scifi and visualization. These include, but are not limited to:

    \begin{itemize}
    \setlength{\itemsep}{0pt}
        \item Posters
        \item Demo proposals or show-and-tell artifacts, including design breakdowns and pipelines
        \item Lightning talks
        \item Two-page position statements
        \item Curated example submissions, where contributors propose a \scifi scene or game moment and explain its visualization relevance
    \end{itemize}
    
    We will organize a reviewing process for the submitted work based on the following procedure:
    \begin{enumerate}
        \item Submissions will be reviewed by the organizers based on their relevance to cross-fertilization, clarity of example, perspective diversity and the potential to spark discussion. If conflicts of interest exist, the organizers will recruit external reviews with the corresponding expertise.
        \item We aim for a balanced program across research and practice, and across media types (e.g., films, TV shows or games). 
        \item If submissions exceed capacity, we prioritize variety and underrepresented perspectives, including early career contributions and practitioners new to VIS.
    \end{enumerate}
    
    We have tentatively chosen the following dates to allow time for outreach. All deadlines are intended as Anywhere on Earth (AoE).
    
    \begin{enumerate}
        \item Call for participation released: April 29, 2026. 
        \item Submission deadline: July 30, 2026. 
        \item Author notification: September 15, 2026. We will ensure this remains before the IEEE VIS early registration deadline, once that date is announced.
        \item Camera-ready deadline: September 21, 2026.
    \end{enumerate}
    
    Direct outreach begins immediately after workshop acceptance, including targeted invitations to practitioners, while keeping the workshop open to all and not invitation-only. We will provide a Discord server for bringing participants together ahead of, during, and after the conference. The general promotion channels will be visualization mailing lists, social media, and practitioner communities such as game development and interface design
    groups.

\section{Organizer Details}
    \textbf{Ulrik Günther} is a Project Lead at the Helmholtz-Zentrum Dresden-Rossendorf e.V., Dresden, Germany.
    Ulrik works at the intersection of biology, visualisation, and human-computer interaction. He has supported the organisation of multiple scientific workshops and loves everything \scifi and XR.
    Website: \texttt{\href{https://ulrik.is/writing}{ulrik.is/writing}}.

    \textbf{Julián Méndez} is a PhD student at the Interactive Media Lab Dresden, TU Dresden, Germany. He develops visual explanations, is in love with the power of (data-driven) storytelling~\cite{MLRBD-2025-ImmDataStoriesReview}, and is a huge fan of messed-up \scifi.
    Website: \texttt{\href{https://imld.de/~mendez}{imld.de/$\sim$mendez}}.
    
    \textbf{Gabriela Molina Le\'{o}n} is a Postdoctoral Research Fellow at Aarhus University, Denmark, with a focus on collaborative visual analytics. She has co-authored several publications on the design and development of multimodal systems to support data experts in performing visual data exploration \cite{molina2022mobile} and collaborative sensemaking \cite{leon25wall}.
    She has co-organized workshops at CSCW 2021, CLIHC 2021, and IEEE VIS 2025.
    Website: \texttt{\href{https://gmleon.github.io/}{gmleon.github.io}}.
        
    \textbf{Samuel Pantze} is a PhD student in computer science at the Center for Advanced Systems Understanding (CASUS), Görlitz, Germany, with a focus on human-in-the-loop interactions in virtual reality and is a 3D artist with a love for \scifi tropes. He has published before on immersive HCI in biological data annotation~\cite{11298768}. 
    Website: \texttt{\href{https://artsnscience.eu}{artsnscience.eu}}.
    
    \textbf{Mario Romero} is a Senior Associate Professor in Immersive Visualization at Linköping University, Sweden. His current research focuses on the role that mixed reality visualization interfaces enhanced with conversational AI agents play in supporting human tasks ~\cite{  bagherifard2025enhancing, iop2023extended, vasiliu2025towards}. Mario has seen, read, and listened to thousands of science fiction stories and strongly suspects that they have had a pervasive influence on his research.
    
    \textbf{Abdulhaq Adetunji Salako} is a PhD Student at the Institute of Visual and Analytic Computing, University of Rostock, Germany. His work focuses on interactive visual metamorphosis for multi-view data exploration \cite{Salako25AnimTransNLPC}, investigating how animated transitions can support understanding across heterogeneous visualization representations.
    Website: \texttt{\href{https://teejay13.github.io/my-profile-site/}{teejay13.github.io/my-profile-site}}.
    
    \textbf{Annalena Ulschmid} is a PhD student at TU Wien, Austria. Her research focuses on guided optimization for re-rendering and re-simulation in interactive systems \cite{Ulschmid2025-cx}.
    Website:
    \texttt{\href{https://www.cg.tuwien.ac.at/staff/AnnalenaUlschmid}{cg.tuwien.ac.at/staff/AnnalenaUlschmid}}.

\acknowledgments{ 
JM was supported by Deutsche Forschungsgemeinschaft (DFG) grant 389792660 as part of TRR~248 -- CPEC (see \url{https://cpec.science}).
AAS was supported by DFG grant \href{https://gepris.dfg.de/gepris/projekt/514630063}{[514630063]}.
GML by the Villum Investigator grant VL-54492 by Villum Fonden.
SP was partially funded by the Center for Advanced Systems Understanding (CASUS) which is financed by Germany’s Federal Ministry of Research, Technology and Space (BMFTR) and by the Saxon Ministry for Science, Culture, and Tourism (SMWK) with tax funds based on the budget approved by the Saxon State Parliament. 
AU was supported by FWF project F77 (SFB \textit{Advanced Computational Design} SP4).

Any opinions, findings, and conclusions expressed in this material are those of the authors and do not necessarily reflect the views of the funding agencies.
}

\bibliographystyle{abbrv-doi-hyperref}

\bibliography{template}
\end{document}